# Modular Autonomous Vehicle in Heterogeneous Traffic Flow: Modeling, Simulation, and Implication

**Lanhang Ye**
Researcher
Institute of Materials and Systems for Sustainability, Nagoya University
Nagoya, Japan 464-8603
Email: ye.lanhang.n2@f.mail.nagoya-u.ac.jp
ORCiD: 0000-0002-4821-1072

**Toshiyuki Yamamoto**
Professor
Institute of Materials and Systems for Sustainability, Nagoya University
Nagoya, Japan 464-8603
Email: yamamoto@civil.nagoya-u.ac.jp
ORCiD: 0000-0002-7540-5040

## ABSTRACT

Modular autonomous vehicles (MAVs) represent a groundbreaking concept that integrates modularity into the ongoing development of autonomous vehicles. This innovative design introduces unique features to traffic flow, allowing multiple modules to seamlessly join together and operate collectively. To understand the traffic flow characteristics involving these vehicles and their collective operations, this study established a modeling framework specifically designed to simulate their behavior within traffic flow. The mixed traffic flow, incorporating arbitrarily formed trains of various modular sizes, is modeled and studied. Simulations are conducted under varying levels of traffic demand and penetration rates to examine the traffic flow dynamics in the presence of these vehicles and their operations. The microscopic trajectories, MAV train compositions, and macroscopic fundamental diagrams of the mixed traffic flow are analyzed. The simulation findings indicate that integrating MAVs and their collective operations can substantially enhance capacity, with the extent of improvement depending on the penetration rate in mixed traffic flow. Notably, the capacity nearly doubles when the penetration rate exceeds 75%. Furthermore, their presence significantly influences and regulates the free-flow speed of the mixed traffic. Particularly, when variations in operational speed limits exist between the MAVs and the background traffic, the mixed traffic adjusts to the operating velocity of these vehicles. This study provides insights into potential future traffic flow systems incorporating emerging MAV technologies.

## INTRODUCTION

As technology continues to advance, the automotive industry is witnessing a paradigm shift towards autonomous vehicles. One fascinating concept on the horizon is the development of modular autonomous vehicles (MAVs), which have the potential to revolutionize traffic flow. This innovative approach combines modularity and autonomy, allowing vehicles to physically connect and detach as needed. This adaptability can reduce aerodynamic drag and energy consumption when traveling on highways in a connected modular train. Once off the highway, the modules can separate to navigate urban areas and reach their destinations more flexibly. A well-known prototype of this innovation is being developed by NEXT Future Mobility, whose modular(https://www.next-future-mobility.com/) electric bus can physically connect and detach, allowing for individual or connected operation. This feature offers unparalleled flexibility, scalability, and customization based on travel demands, transforming urban mobility, last-mile logistics, and overall transportation efficiency.

However, despite their considerable potential, the deployment of such innovations in a practical settings is challenging and requires comprehensive investigation and resolution. One particular concern pertains to how the integration of these vehicles and their subsequent collective operations will affect the existing traffic system and its potential impact on traffic flow. Introducing modularity into the current traffic infrastructure might necessitate a phased transition, involving the simultaneous presence of human-operated, autonomous, and modular vehicles. This evolution could introduce complexities and potential disruptions in traffic patterns, which must be carefully monitored and managed.

For instance, the implementation of modular operation raises legal and regulatory issues. Policymakers must formulate comprehensive guidelines and standards to ensure the safe and responsible operation of these vehicles on public roads. These regulations should address vital aspects such as defining the maximum size of modular units in a coupled formation and establishing conditions for the safe docking and detaching of modules within existing units. While the maximum size of modules in a single coupled train might not pose a technical challenge, it could directly affect traffic flow dynamics and result in potential safety problems. Striking the right balance is of paramount importance, as smaller coupled units may not fully exploit the advantages of scalability and modularity, while excessively long coupled trains could lead to congestion and pose challenges for maneuvering on busy roads.

Policymakers, in collaboration with industry experts and researchers, should conduct comprehensive simulations and safety assessments to determine an optimal coupled module size that maximizes benefits while minimizing potential risks. Another critical aspect of regulation involves specifying the conditions under which individual modules are permitted to dock and detach from existing formations. Developing clear and standardized protocols for these actions is crucial for maintaining traffic flow, ensuring safety, and preventing potential disruptions to the transportation system.

To tackle the challenges of implementing this innovation into the current traffic system, developing specialized models that can examine traffic flow dynamics is crucial, while also considering the incorporation of such vehicles and their operations within mixed traffic flow. Conventional traffic flow models and simulation platforms, such as widely used commercial software like VISSIM, cannot adequately model these innovative concepts and unprecedented operations. To address this issue, this study introduces a customized traffic flow model to simulate modular vehicle operations within mixed traffic environments and to investigate concerns related to traffic-flow dynamics in the presence of these vehicles. The established model employs a discrete mode-based approach, enabling modules to operate independently, form coupled trains of various sizes, and efficiently join or detach from existing formations according to specific operational rules. Integrated discrete switching logic ensures smooth transitions between different modes, with clearly defined conditions determining when modules can join or detach from coupled units. This allows them to reach their destinations collectively or individually in a more flexible manner.

The objective of this research is to investigate the dynamics of traffic flow involving MAVs and their operations within a mixed traffic environment. Comprehensive simulations and sensitivity analyses are employed to evaluate and understand this potential future traffic system. This initiative begins with the introduction of a specialized traffic flow model, encompassing both microscopic and macroscopic characteristics. Key factors such as traffic demand levels and the MAV penetration rate within the traffic system are also considered. The study aims to provide insights into the integration of MAVs into current transportation infrastructure, facilitating a safe, efficient, and seamless transition. Additionally, this study endeavors to comprehend the implications of traffic flow theory in the context of the emerging mobility era.

As a cutting-edge concept, MAVs represent a substantial advancement beyond the current wave of autonomous vehicle technology by enhancing autonomy through modularity. This distinctive feature sets MAVs apart from conventional traffic configurations, allowing these vehicles to function either independently or collectively and adapt dynamically to evolving traffic conditions, routes, and destinations. This capability holds the potential to improve traffic efficiency, safety, and sustainability in unprecedented ways. In contrast, conventional traffic flow primarily involves individual vehicles operating independently, lacking such modular capabilities. These disparities introduce new challenges that conventional traffic flow models and simulation platforms may find difficult to address. Consequently, researchers are increasingly focusing on these emerging challenges from various perspectives.

Modular vehicle technology is widely expected to have considerable potential in public transit. Numerous existing studies have been conducted in this field, addressing topics such as system design, scheduling, and demand management for novel transit systems using MAVs. For instance, Chen et al. developed discrete and continuous modeling methods for designing the operation of a shuttle system using modular vehicles in the context of oversaturated traffic, representing pioneering efforts in introducing modularity into urban mass-transit systems (*1*, *2*). Zhang et al. developed a mathematical model for modular transit within a time-expanded network and examined the performance of modular buses under various configurations. Their research suggests that this system can serve not only as a local transportation option but also cater to long-distance trips by providing efficient and convenient connections to main modules, resulting in more efficient vehicle utilization and serving a larger number of passengers compared to conventional door-to-door shuttle services (*3*).

Wu et al. introduced a modular, adaptive, and autonomous transit system that can adapt to dynamic traffic demands without being limited to fixed routes and timetables through transfer operations, envisioning a futuristic public transit system employing such innovative technologies (*4*). Chen and Li conducted a study on corridor systems involving MAVs, which allows for station-wise docking/undocking. Their research explored the potential of MAV-based station-wise docking design in public transport, highlighting benefits such as reduced system costs and increased efficiency (*5*). Chen et al. further introduced a continuous model to offer near-optimal solutions for designing MAV-based transit corridors, ensuring efficient operation of the proposed system (*6*).

Lin et al. conducted a forward-looking study covering a wide range of perspectives on the challenges and opportunities for the future of the autonomous modular mobility paradigm, considering transit service and last-mile delivery service across various phases, including planning, infrastructure design, operation, and business models (*7*). Li and Li introduced trajectory planning and optimization solutions for modules during docking and split operation processes, improving riding comfort and fuel efficiency and facilitating efficient operations of modules between long coupled trains and multiple shorter ones (*8*, *9*). Han et al. introduced a hierarchical docking planning model using Nonlinear Model Predictive Control to address the challenges of trajectory planning for the physical docking of modular buses. Their approach includes the design of objective functions, docking constraints, and avoidance constraints, ensuring optimal performance across diverse docking scenarios (*10*).

These studies collectively demonstrate the potential of modularity in the context of mass transit systems, underscoring its ability to enhance efficiency, flexibility, and scalability in public transportation.

In the field of traffic flow modeling, the vision of innovative systems, such as MAVs, is an emerging topic that has not yet been well studied. Conventional traffic flow models predominantly focus on individual-level operations, with most emphasizing the car-following process. Although several pioneering efforts have been made, particularly in trajectory planning and optimization during docking and split operation processes of MAVs (*8*, *9*,*10*), these models are unable to evaluate operations at a system level or delve into the traffic flow dynamics encompassing the operations of such vehicles. This gap in research could be attributed to the incapability of existing traffic flow models to simulate MAV operations, including the docking and detachment processes involving longer coupled modules, smaller ones, or individual modules operating within mixed traffic.

To the best of the authors' knowledge, no study has investigated the collective operations of MAVs and evaluated their performance on traffic flow at the system level. Recognizing this research gap, this study aims to develop a tailored cellular automata-based model capable of integrating this emerging innovation and modeling their unique operations in mixed traffic flow. The objective is to gain a deeper understanding of this emerging technology and envision a future scenario involving these vehicles through comprehensive simulation and analysis.

## METHODOLOGY

The cellular automata (CA) model is a discrete, rule-based model that finds widespread application in microscopic traffic flow modeling. In this study, a modeling framework specifically designed for incorporating MAV operations within mixed traffic flows is developed. The model comprises two main components: a background flow model designed for simulating conventional vehicles and a mode-dependent model intended for simulating MAV operations within mixed traffic flow. Two types of vehicles are included: conventional vehicles and MAVs. The time step in the model represents 1 s, and velocity and acceleration rates are measured in m/s and m/s², respectively. The following rules are defined for each vehicle per second.

### Two-lane TSM for simulating background traffic

To model conventional vehicles, the same rules as the two-state safe-speed model (TSM) are applied. The TSM model, first proposed by Tian et al., can reproduce complex traffic flow characteristics of conventional traffic flow, including metastable states, traffic oscillations, and phase transitions (*11*). This study further extends this model to a two-lane configuration by applying a classical lane-changing model to simulate background traffic, which was first introduced in our previous work (*12*). The model uses a refined cell length of 0.5 m and follows the updating procedure outlined below, the frequently used abbreviations and notations are listed in Appendix A:

*Deterministic speed update*

$$v'_{\text{det}} = \min(v + a, v_{\max}, d_{\text{anti}}, v_{\text{safe}}) \tag{1}$$

In this step, the speed of the target vehicle at the next time step is determined as the minimum of the following: the speed under acceleration, the maximum speed, the anticipated space gap, and the safe speed.

*Stochastic deceleration*

$$v' = \begin{cases} \max(v'_{\text{det}} - b_{\text{rand}}, 0) & \text{with probability } p \\ v'_{\text{det}} & \text{otherwise} \end{cases} \tag{2}$$

The stochastic deceleration step accounts for potential speed reductions of the target vehicle due to random factors encountered during driving.

*Position update*

$$x' = x + v' \tag{3}$$

The position update indicates that the target vehicle will move forward by a distance corresponding to the updated speed within 1 s of the next time step. Here, $v$ ($v'$) and $x$ ($x'$) denote the speed and position at the current and subsequent time steps, respectively. $a$ and $v_{\max}$ are the acceleration rate and maximum velocity of the vehicle, respectively. $b_{\text{rand}}$ denotes the randomization-deceleration rate. $d_{\text{anti}}$ denotes the anticipated space gap, $v_{\text{safe}}$ denotes the safe speed defined in the Gipps model (*13*). $d_{\text{anti}}$ and $v_{\text{safe}}$ are defined as follows

$$d_{\text{anti}} = d + \max\left(v_{\text{anti}} - g_{\text{safety}}, 0\right) \tag{4}$$

$$v_{\text{safe}} = \left[-b_{\max} + \sqrt{b_{\max}^2 + v_l^2 + 2b_{\max}d}\right] \tag{5}$$

These equations assume: (i) a reaction time of 1 s (which corresponds to the time step of the CA model) and (ii) no acceleration during the current time step.
$d = x_l - x - L_{\text{veh}}$ is the real space gap. $L_{\text{veh}}$ denotes the length of the vehicle.
$v_{\text{anti}}$ denotes the expected velocity of the preceding vehicle.

$$v_{\text{anti}} = \min(d_l, v_l + a, v_{\max}) \tag{6}$$

$x_l$, $d_l$, and $v_l$ denote the position, real space gap, and speed of the preceding vehicle, respectively. $g_{\text{safety}}$ denotes a safety parameter that helps avoid accidents considering the limitation of human perception, with the constraint $g_{\text{safety}} \geq b_{\text{rand}}$. $b_{\max}$ denotes the maximum deceleration rate. The round function $[x]$ helps return the integer nearest to $x$.
The randomization deceleration $b_{\text{rand}}$ and stochastic deceleration probability $p$ are specifically defined as follows:

$$b_{\text{rand}} = \begin{cases} a & \text{if } v < b_{\text{defense}} + \left\lfloor \dfrac{d_{\text{anti}}}{T} \right\rfloor \\ b_{\text{defense}} & \text{otherwise} \end{cases} \tag{7}$$

$$p = \begin{cases} p_{\text{b}} & \text{if } v = 0 \\ p_{\text{c}} & \text{else if } v \leq \dfrac{d_{\text{anti}}}{T} \\ p_{\text{defense}} & \text{otherwise} \end{cases} \tag{8}$$

where $p_{\text{defense}} = p_{\text{c}} + \frac{p_{\text{a}}}{1+e^{\alpha(v_{\text{c}}-v)}}$ is a logistic function used to define the randomization probability. In the function $b_{\text{rand}}$, two distinct randomization-deceleration values are used to describe differences in driving behavior under two different states: the defensive and normal states. $\lfloor x \rfloor$ represents the floor function, which returns the largest integer no greater than $x$. $b_{\text{defense}}$ is the randomization-deceleration rate for the defensive state. $T$ is the effective safe time gap. The defensive state is activated if $v > d_{\text{anti}}/T$. In the normal state, the randomization-deceleration rate is equal to $a$.

*Symmetry lane-changing*

Incentive criteria:

$$d(i,t) < \min\{v + a, v_{\max}\} \tag{9}$$

and

$$d(i,t)_{\text{other}} > \min\{v + a, v_{\max}\} \tag{10}$$

Safety criterion:

$$d(i,t)_{\text{back}} > v_{\max} \tag{11}$$

The incentive criteria indicate that the space ahead of the object vehicle $i$ is insufficient for traveling at a higher velocity, and that the driving conditions on the target lane are better than those on the current lane. The safety criterion ensures that, when changing lanes, the vehicle immediately behind the object vehicle on the target lane will not crash into the object vehicle after it has moved. When both conditions are met simultaneously, the object vehicle will change to the target lane with a lane-changing probability $P_{\mathrm{lc}}$.

**Modeling MAVs operation in mixed traffic**

Compared to conventional vehicles, a unique characteristic of MAV operations is the docking and detaching processes. Docking refers to the process of vehicles joining an existing MAV train or forming a new one, with the preceding module operating independently, while detaching involves a modular unit separating from an existing MAV train to reach its destination. To integrate these operations within mixed traffic flow, a mode-dependent approach has been implemented. This approach includes different car-following rules for each operation mode and uses discrete switching logic to manage transitions between modes throughout the processes. This innovative approach, introduced in this study, facilitates the incorporation of MAV operations and is a critical component of the mixed flow model.

Three distinct modes of operation for MAVs have been defined: independent moving, docking operation, and collective moving. In the independent moving mode, MAVs operate autonomously and independently, without being part of any coupled module. The docking operation mode involves joining an existing MAV train or forming a new one, where the leading module may operate independently during the docking process. The collective moving mode indicates that MAVs move collectively as a single unit, with either zero intra-platoon spacing or a constant predefined intra-platoon spacing. These modes correspond to different stages in MAV operations and are essential for their efficient and coordinated functioning. Additionally, rules governing the detaching processes have been established and integrated into the model within the context of lane-changing processes. The corresponding rules for each operation mode are provided as follows:

*Independent moving mode*

In the independent moving mode, the speed update rules for individual MAV modules follow those of conventional vehicles as outlined in the preceding section. The primary distinction is the omission of stochastic deceleration, as MAV modules operate autonomously, providing more precise control and enhanced stability.

A lower speed limit, denoted as $v'_{\max}$ , is imposed for MAV operations in both the independent moving mode and the collective moving mode. This limit corresponds to the actual operational speed limit of the NEXT modular vehicle.

**Table 1** presents illustrations of mixed traffic flow consisting of regular vehicles and MAVs. The illustrations depict a 250-m road segment over several consecutive time steps, with regular vehicles shown in black and MAVs in independent moving mode shown in blue.

**TABLE 1 MAV Modules Operating Independently in Mixed Traffic Flow**

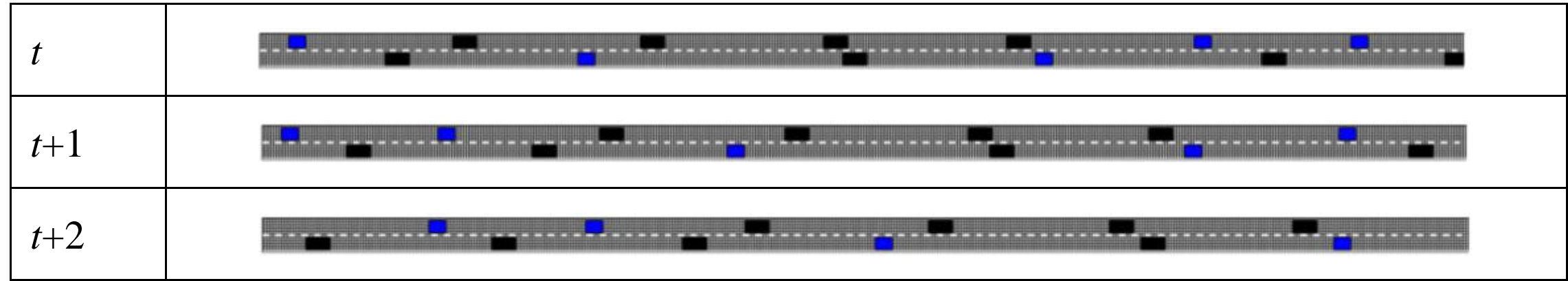

| $t$ | |
|---|---|
| $t$+1 | |
| $t$+2 | |

*Docking operation mode*

The docking mode involves the sequential coupling of two modules to form a MAV train or the integration of a new module into an existing MAV train. During this process, it is crucial to reduce the distance between the target module and its designated partner and minimize the speed difference between them to ensure seamless docking. This can be achieved through a catch-up strategy, where the following module accelerates to match the speed and position of the preceding module or MAV train.

The conditions for transitioning a module from the independent moving mode to the docking operation mode are as follows:

i. The preceding vehicle is a MAV.

ii. The preceding MAV train is smaller in size than the maximum number of modular units allowed $L_{\max}$.

When either of these conditions is met, a docking event is triggered, causing the following module to switch to the docking operation mode. In this mode, the vehicle will adjust its speed according to the following rules:

$$v' = \min\left(v + a_p', v_{\max}, d - d_{\text{intra}}\right) \tag{12}$$

where $a_p'$ is the acceleration rate during the docking process. $v_{\max}$ represents the maximum speed of the MAV module in docking mode, and the condition $v_{\max} > v_{\max}'$ ensures the proper execution of the docking process in a free-flow state. $d_{\text{intra}}$ denotes the spacing in docked mode, which is a predefined fixed variable in the simulation settings.

The docking process will be halted under two conditions: if another vehicle enters the gap between the target MAV and the module in front, or if neither of the previously mentioned conditions is satisfied. In such cases, the module will revert to independent moving mode.

**Table 2** presents illustrations of MAV docking operations. In these illustrations, MAVs in docking operation mode are shown in green, while MAV trains are shown in red.

**TABLE 2 MAV Docking Operations in Mixed Traffic flow**

| | |
|---|---|
| *t* | |
| *t*+1 | |
| *t*+2 | |
| *t*+3 | |

*Collective moving mode*

Collective moving means that docked modules use the same driving parameters as their coupled module leader and maintain constant spacing with each other.

If $a_p' = 0$ and $d = d_{\text{intra}}$, the vehicle will switch from docking mode to collective moving mode. During the deterministic speed update step, the following vehicles in the MAV train will be updated to the same speed as the coupled module leader $v_{\text{det}}^{\text{leader}}$:

$$v' = v_{\text{det}}^{\text{leader}} \tag{13}$$

For the MAV train leader, the same deterministic speed update applies as for conventional vehicles, with the stochastic deceleration step omitted.

**Table 3** presents illustrations of MAV trains operating in collective moving mode.

**TABLE 3 MAV Train Moving Collectively in Mixed Traffic flow**

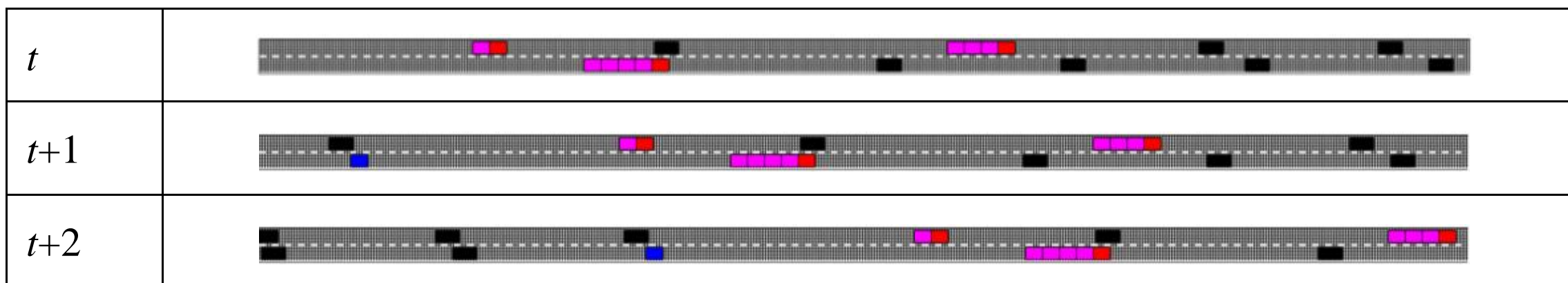

*Lane-changing and detachment*

The lane-changing behavior of MAVs is defined under various operation modes, each tailored to different initiatives and requirements for achieving effective MAV operation in mixed traffic flow.

In the independent moving mode, MAV lane-changing behavior introduces a new incentive criterion related to the formation or joining of an existing MAV train in an adjacent lane. This additional criterion is incorporated alongside the existing incentive criteria for background traffic and is as follows:

Incentive criteria:

i.

$$d(i,t) < \min\{v + a, v_{\max}\} \tag{14}$$

and

$$d(i,t)_{\text{other}} > \min\{v + a, v_{\max}\} \tag{15}$$

ii. If the preceding vehicle on the current lane is a conventional vehicle, and the preceding vehicle on the target lane is an MAV.

Safety criterion:

$$d(i,t)_{\text{back}} > v_{\max} \tag{16}$$

In the docking operation mode, lane-changing behavior is prohibited. This restriction is because, in the independent moving mode, the specific initiative for the MAV is focused on catching up with the target MAV or MAV train ahead.

In the collective moving mode, lane-changing involves the process of detaching a module from a connected MAV train. The parameter $P_{\text{d}}$ denotes the probability of a docked module leaving the current MAV train. This departure may occur for reasons such as deviating from the train's route or exiting the highway at an interchange. This parameter reflects the potential for platoons to reassemble in the simulated system. The detaching process from the MAV train resembles the lane-changing process and is influenced by surrounding traffic conditions, particularly in congested situations. The rules governing this process are as follows:

Incentive criterion:

$$P_{\text{rand}} < P_{\text{d}} \tag{17}$$

Safety criterion:

$$d(i,t)_{\text{back}} > v_{\max} \tag{18}$$

The parameter $P_{\text{rand}} \in (0,1)$ is a randomly generated number. If the specified conditions are met, the module will detach from the MAV train and change its current lane. If the module serves as the MAV train leader and more than two coupled modules following it are present in the remaining MAV train, the module immediately succeeding it will assume the role of the new train leader. However, if only one module is remaining, it will transition to independent moving mode, resulting in the complete detachment of the original train. For the detachment of a following module, the outcome depends on its position within the original MAV train. The original MAV train may be reorganized into two sub-trains, or a portion of the original train may detach, depending on the specific circumstances. Illustrations of module detachment are presented in **Table 4**.

**TABLE 4 MAV Module Detachment in Mixed Traffic flow**

| | |
|---|---|
| *t* | |
| *t*+1 | |
| *t*+2 | |
| *t*+3 | |

**Basic assumptions**

Based on the characteristics of each stage in MAV operations, a corresponding modeling method for mixed traffic flow was established. This approach encompasses the entire process of MAV operations, including docking, collective moving, and detaching. The developed model is a two-lane model with symmetrical lane-changing rules applied to all vehicles, and there is no preference for either lane of the simulated two-lane highway. Note that the lane-changing behavior of MAV trains was not considered in this study. The model is built on the following fundamental assumptions:

1. MAVs are assumed to exhibit driving performance similar to conventional vehicles, with the primary difference being reduced randomness during driving. The focus is on integrating MAV operations within mixed traffic flow, and any disparities in driving performance between MAVs and conventional vehicles are considered outside the scope of this study.

2. It is assumed that, under ideal communication and operational conditions, MAVs will maintain constant spacing within the MAV train and adhere strictly to the defined rules. Issues related to control and stability within the MAV train are not addressed in this study. When the specified spacing is set to zero, it indicates a physical connection between a module and its coupled module or MAV train.

**Simulation setup**

In the CA model, road segments are discretized into cells, with each cell measuring 0.5 m in length and each time step representing 1 s. The states of each cell at any given time step are binary, indicating whether the cell is occupied or empty. A typical vehicle occupies a row of 10 consecutive cells, which includes its physical length of 5 m plus the minimum clearance required in stationary queuing conditions. Simulations were conducted on a 10-km two-lane loop road segment, ensuring a constant vehicle density, which is used as a variable to evaluate MAV operations within mixed traffic flow.

The initial state involves two types of vehicles randomly distributed across the road segment. Each simulation runs for 12,000 time steps, with MAV docking operations starting after 5,000 time steps. To exclude transition effects, the results from the initial 10,000 time steps are discarded. Four scenarios are simulated, each with different proportions of MAVs denoted by $P_{\mathrm{mav}}$: 0%, 25%, 50%, and 75%. Simulations are conducted at various traffic demand levels, covering the full range of density conditions.

**Table 5** and **Table 6** list the parameters for the TSM model and those used to model MAV operations, respectively.

**Table 5 Parameters of Two-lane TSM Model**

| **Parameters** | $L_{\mathrm{cell}}$ | $L_{\mathrm{veh}}$ | $v_{\mathrm{max}}$ | $T$ | $a$ | $b_{\mathrm{max}}$ | $b_{\mathrm{defense}}$ | $P_{\mathrm{a}}$ | $P_{\mathrm{b}}$ | $P_{\mathrm{c}}$ | $g_{\mathrm{safety}}$ | $v_{\mathrm{c}}$ | $\alpha$ | $P_{\mathrm{lc}}$ |
|---|---|---|---|---|---|---|---|---|---|---|---|---|---|---|
| **Units** | m | $L_{\mathrm{cell}}$ | m/s | s | m/s$^2$ | m/s$^2$ | m/s$^2$ | - | - | - | $L_{\mathrm{cell}}$ | $L_{\mathrm{cell}}$/s | s/$L_{\mathrm{cell}}$ | - |
| **Values** | 0.5 | 10 | 33 | 1.8 | 1 | −3 | 1 | 0.85 | 0.52 | 0.1 | 20 | 30 | 10 | 0.2 |

**Table 6 Parameters for Modeling MAV**

| **Parameters** | $L_{\mathrm{mav}}$ | $v'_{\mathrm{max}}$ | $a'_{\mathrm{p}}$ | $d_{\mathrm{intra}}$ | $P_{\mathrm{d}}$ | $L_{\mathrm{max}}$ |
|---|---|---|---|---|---|---|
| **Units** | $L_{\mathrm{cell}}$ | m/s | m/s$^2$ | m | - | veh |
| **Values** | 7 | 30.5 | 1 | 0 | 0.2 | 5 |

$L_{\mathrm{cell}}$ represents the length of each cell of the road segment, which is 0.5 m in this study. $L_{\mathrm{veh}}$ denotes the length of background vehicle and is equal to 5 m. The maximum speed limit $v_{\mathrm{max}}$ is equal to 33 m/s. For MAV, the operational maximum speed limit $v'_{\mathrm{max}}$ is 30.5 m/s, except for that in docking mode, in which the maximum speed limit $v_{\mathrm{max}}$ is applied. $L_{\mathrm{max}}$ represents the maximum modular units' size in a MAV-train. $d_{\mathrm{intra}}$ represents spacing between individual MAV modules within MAV train. The other parameters used in the two-lane TSM model adhere to the settings originally proposed by Tian et al. (*11*). The speed limits for vehicles and the maximum number of modular units in MAV operations are based on the actual configuration of the NEXT Modular Vehicle. Specifically, the maximum speed of the MAV is set at 110 km/h, and the maximum number of modular units, $L_{\mathrm{max}}$, in an MAV train is 5.

## RESULTS AND DISCUSSION

The result from an individual run is first presented to demonstrate the simulation. The size distributions of arbitrarily formed MAV trains in mixed traffic flow from 9 simulations under various parameters were analyzed. Finally, extensive simulations were conducted, and fundamental diagrams were generated for two comparative scenarios, and sensitivity analyses were performed on MAV penetration rates in mixed traffic flow, comparing these results with a base scenario without MAVs. The primary difference between the two scenarios is whether MAVs are permitted to form trains and move collectively within the mixed traffic flow.

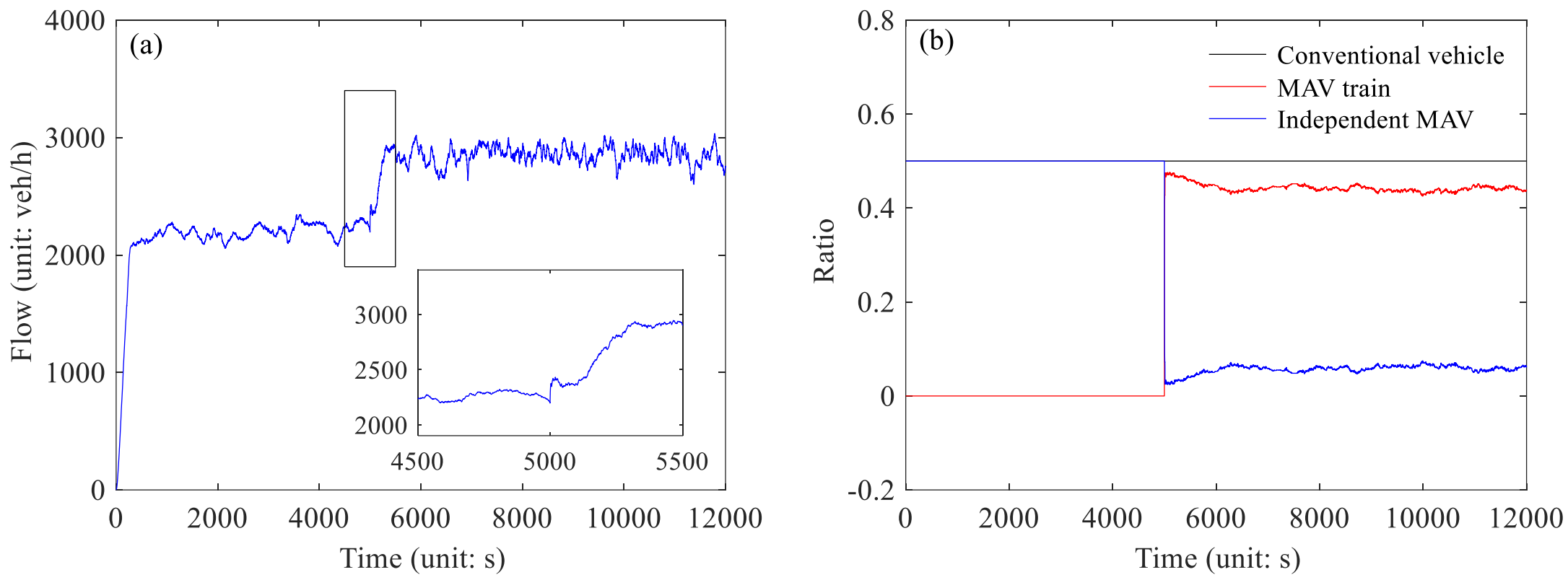

**Figure 1 Results of individual simulation: (a) Flow Time Series, (b) Ratios of vehicle compositions**

Figure 1 illustrates the flow time relation and ratios of vehicle compositions in mixed traffic flow through the simulation with specific penetration rates ($P_{\mathrm{mav}}$) of 50% and traffic demand level of 60 veh/km (medium traffic). In Figure 1(a), a notable surge in the flow rate occurs at the 5000-time step mark, which is when MAVs begin their docking operations and start to operate collectively as coupled MAV trains. In Figure 1(b), corresponding ratios of MAV moving independently and collectively are reveled. The simulation consists of two phases, divided at the 5000-time step mark. Before this point, the simulation produces a mixed traffic flow with conventional vehicles and MAVs moving independently. After this point, the MAVs are allowed to form coupled trains and move collectively. It is essential to note that not all MAV modules will form MAV trains and operate in a collective moving mode; a small fraction of MAV modules will continue to operate independently.

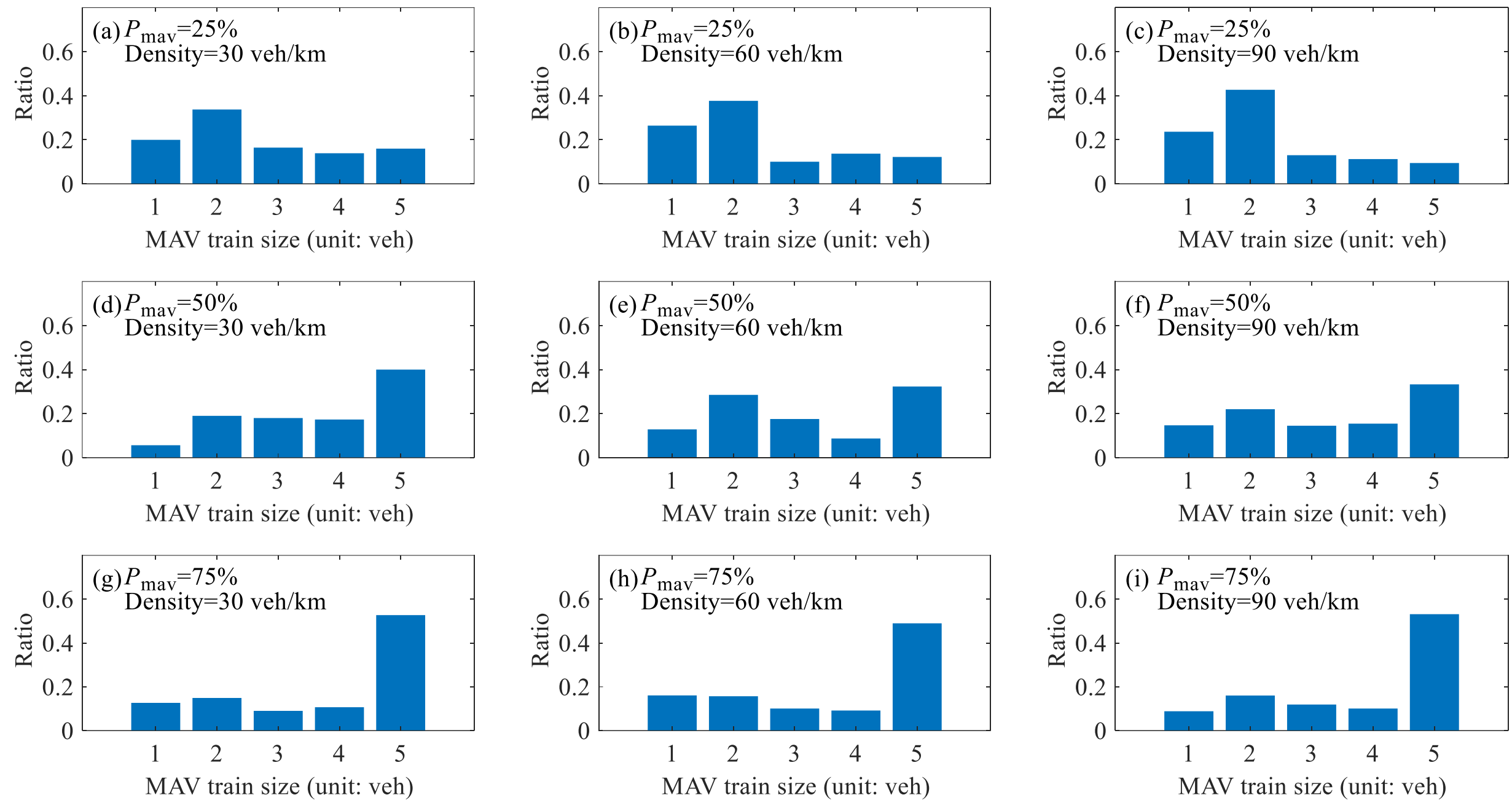

**Figure 2 MAV train size distributions under various penetration rates and traffic demand levels**

Figure 2 illustrates the size distributions of MAV trains under varying penetration rates ($P_{\text{mav}}$) of 25%, 50%, and 75%. Three traffic demand levels are considered: 30 veh/km (light traffic), 60 veh/km (medium traffic), and 90 veh/km (dense traffic). At a $P_{\text{mav}}$ of 25%, as shown in subplots (a), (b), and (c), a significant proportion of the trains consist of only 2 modules. This occurs because, at lower penetration rates, MAV modules are sparsely distributed in the mixed traffic flow, which hampers the formation of larger trains. As $P_{\text{mav}}$ increases, the proportion of MAV trains with larger module sizes also increases, with a notable rise in trains with 5 modules. At a 75% penetration rate, most trains consist of 5 modules, indicating that the penetration rate is a crucial factor influencing MAV train size distribution and operations. The size distribution of MAV trains in mixed traffic is influenced by both traffic demand levels and MAV penetration rates, reflecting the complex dynamics introduced by the integration of modular vehicles. This analysis of module size distributions connects microscopic characteristics with macroscopic fundamental diagram analysis discussed in the subsequent section.

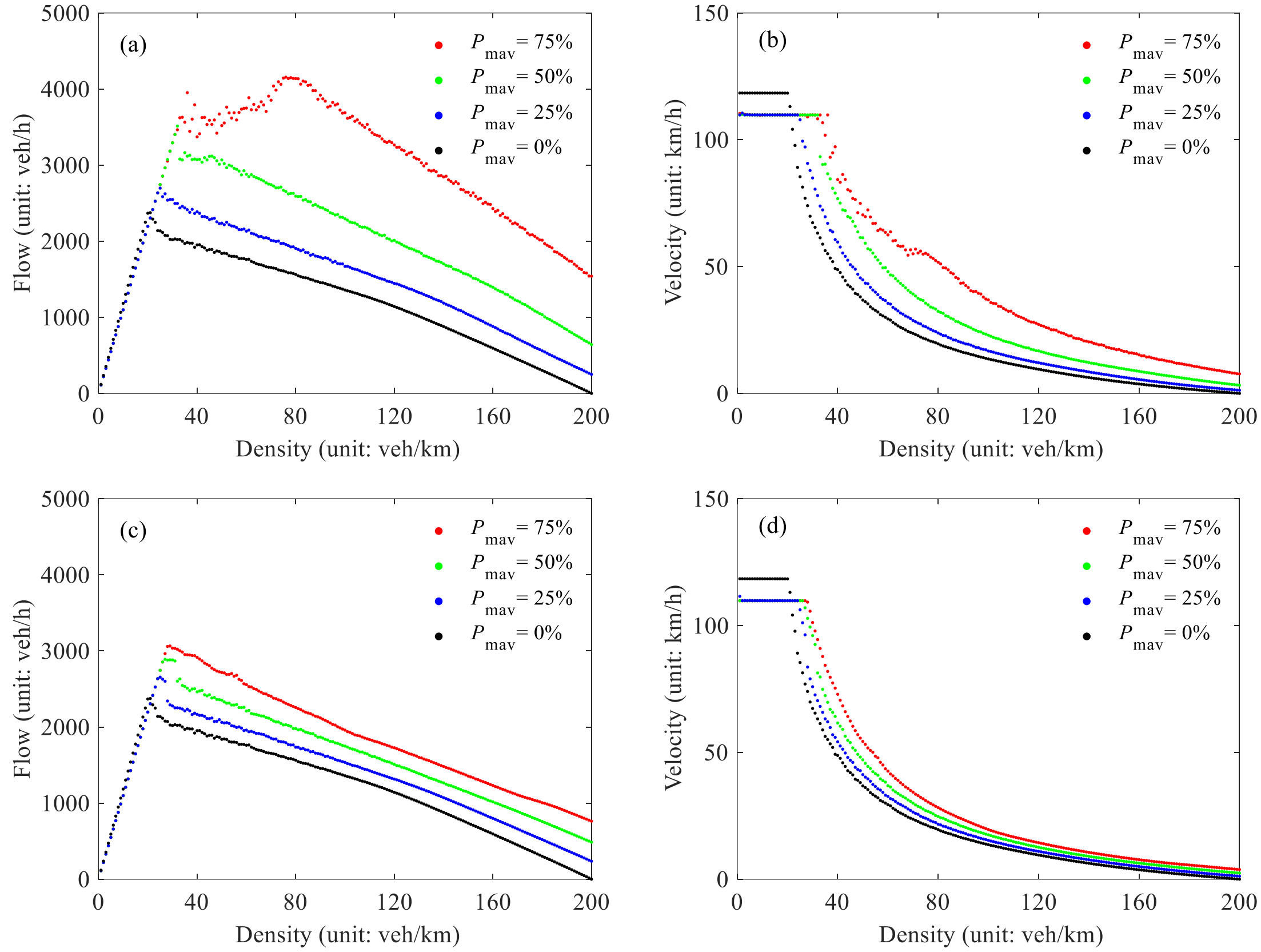

**Figure 3 Fundamental diagrams of scenarios: (a, b) with MAV train operations, and (c, d) without MAV train operations at various MAV penetration rates.**

Figure 3 presents the fundamental diagrams of mixed traffic flow with and without MAV modules. The diagrams compare two scenarios: (a, b) where MAVs form grouped trains and operate collectively, and (c, d) where MAV modules operate independently. Each scenario includes four cases with varying MAV penetration rates: 0%, 25%, 50%, and 75%. The fundamental diagram represents the flow-density relationship for each lane in the simulation.

Figure 3(a) and 3(c) illustrate the relationship between flow and density, while Figure 2(b) and 2(d) depict the relationship between velocity and density. The first scenario features mixed traffic flow consisting of regular vehicles, MAV modules, and arbitrarily formed MAV trains of different sizes, while the second scenario involves regular vehicles and individually operating MAV modules.

In Figure 3(a), the fundamental diagram for the flow–density relationship highlights several distinctions in the presence of MAVs and MAV trains. Notably, capacity increases with the penetration rate of MAVs. This increase results from several factors: the higher number of MAVs (each 3.5 m long compared to 5 m for regular vehicles), reduced randomization in MAV operations, and the collective operation of MAVs as coupled trains. In contrast, Figure 3(c) shows the flow–density relationship without the collective operation of MAVs, capturing only the effects of MAVs' length and reduced randomization.

Figure 3(b) and 3(d) illustrate the velocity-density relationships for both scenarios. When MAVs are integrated into mixed traffic flow, the free-flow speed decreases to match the maximum operational velocity of the MAVs. This reduction is due to the difference in maximum operational velocities between MAVs and regular vehicles, similar to scenarios involving mixed traffic with varying maximum speeds.

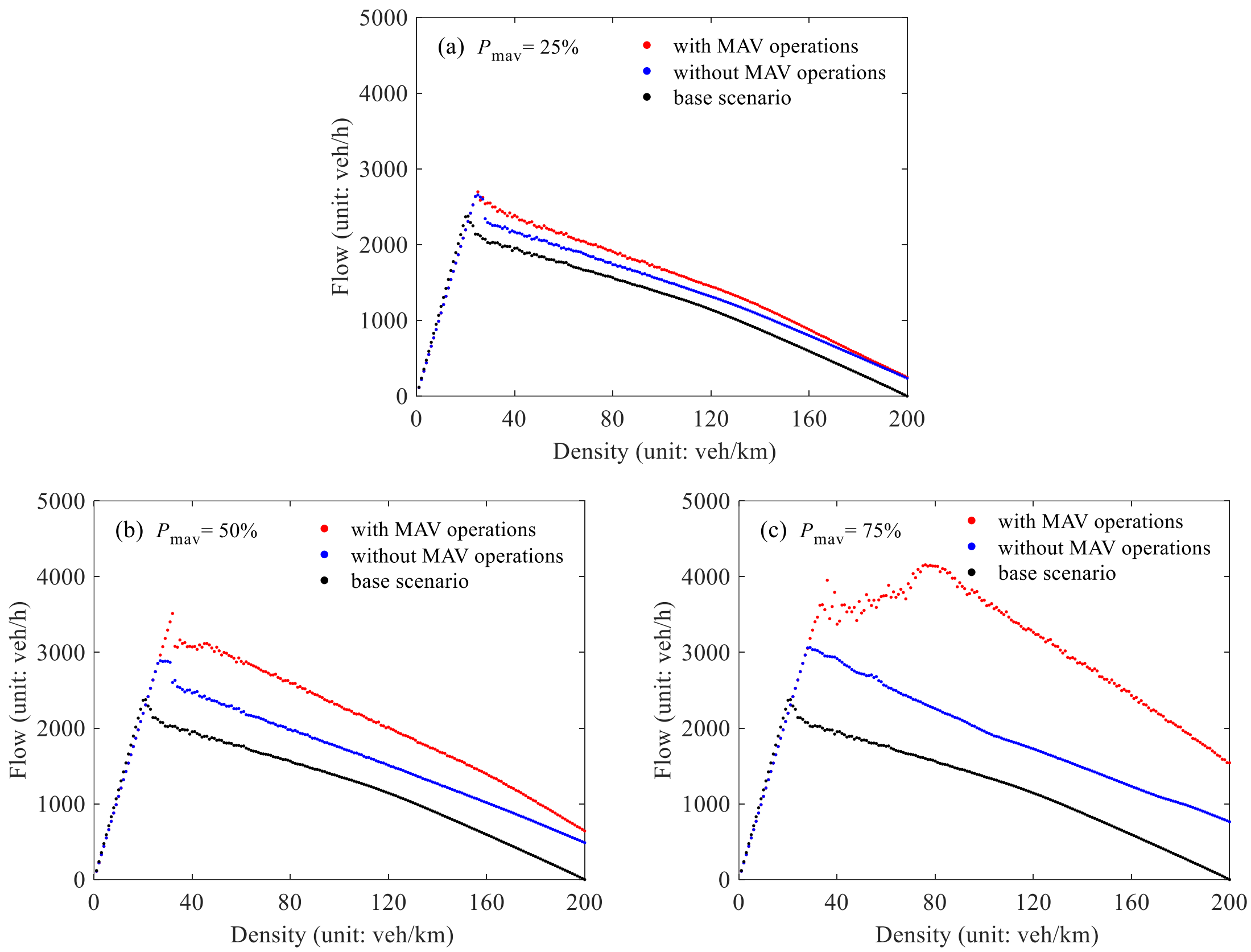

**Figure 4 Fundamental diagrams of comparative cases at various MAV penetration rates.**

Figure 4 presents the flow–density diagrams for three comparative cases with different MAV penetration rates: (1) the base case without MAVs, (2) the case with MAVs operating independently, and (3) the case with MAVs operating collectively. The differences between the flow–density curves in these cases highlight the impact of MAVs and their operational modes. Specifically, the space between the blue and black curves represents the impact of MAVs operating independently, while the space between the red and blue curves shows the impact of MAVs operating collectively. As the MAV penetration rate increases, the influence of MAVs operating collectively becomes more pronounced. This is because a higher MAV penetration rate enhances the likelihood of forming larger MAV trains, thereby significantly improving traffic flow rates through collective operations.

## CONCLUSIONS

This study focused on the modeling and simulation of MAV operations within mixed traffic flow. A two-lane cellular automata-based traffic flow model was developed to facilitate this analysis. Numerical simulations were performed at various MAV penetration rates and different levels of traffic demand. The study examined MAV train size distributions and macroscopic fundamental diagrams across comparative scenarios to evaluate the performance of mixed traffic flow with MAVs and their collective operations. The goal was to provide a comprehensive understanding of traffic flow dynamics with the integration of modular automotive technology.

Simulation results indicate that incorporating MAVs and their collective operations can significantly increase traffic capacity, with capacity nearly doubling when the MAV penetration rate exceeds 75%. However, the inclusion of MAVs and MAV trains also influences and regulates the free-flow speed of mixed traffic, particularly when there are differences in operational speed limits between MAVs and conventional vehicles.

This study developed a microscopic traffic flow model to simulate MAV operations within mixed traffic. The parameters governing MAV operations, including the maximum number of modules in a single MAV train and velocity limits, align with the current configurations of NEXT Future Mobility. Future research will focus on sensitivity analysis of regulatory parameters to provide further insights into MAV operations and evaluate the impact of these parameters on traffic flow performance, addressing both safety and efficiency considerations.

## ACKNOWLEDGMENTS

This work was supported by JST Grant Number JPMJPF2212. The authors would like to thank Mr. Tommaso Gecchelin for generously providing technical information regarding the NEXT Modular Vehicles.

## AUTHOR CONTRIBUTIONS

The authors confirm their contributions to the paper as follows: study conception and design, data collection, analysis and interpretation of results, and draft manuscript preparation: Lanhang Ye; review and editing: Toshiyuki Yamamoto. All authors reviewed the results and approved the final version of the manuscript.

## APPENDIX

### A. List of abbreviations and notations

The following are important abbreviations and notations utilized throughout this paper:

| Notation | Description |
|---|---|
| MAV | Modular autonomous vehicle |
| AV | Autonomous vehicle |
| CA | Cellular automata |
| TSM | Two-State Safe-Speed Model |
| $v'_{\mathrm{det}}$ | Deterministic speed |
| $v$ | Velocity at current time step |
| $a$ | Acceleration rate |
| $v_{\max}$ | Maximum speed limit |
| $d_{\text{anti}}$ | Anticipated space gap |
| $v_{\text{safe}}$ | Safety speed |
| $v'$ | Stochastic speed |
| $b_{\text{rand}}$ | Stochastic deceleration rate |
| $p$ | Stochastic deceleration probability |
| $x\ (x')$ | Position at current time step (subsequent time step) |
| $d$ | Space gap of the subject vehicle |
| $x_l$ | Position of leading vehicle |
| $L_{\text{veh}}$ | Length of vehicle |
| $v_{\text{anti}}$ | Expected velocity of leading vehicle |
| $d_l$ | Space gap of leading vehicle |
| $v_l$ | Velocity of leading vehicle |
| $g_{\text{safety}}$ | Safety parameter |
| $p_{\text{defense}}$ | Randomization probability |
| $d(i,t)$ | Space gap of vehicle *i* at time step *t* |
| $d(i,t)_{\text{other}}$ | Space gap of vehicle *i* at time step *t* on target lane |
| $d(i,t)_{\text{back}}$ | Space gap of vehicle behind vehicle *i* at time step *t* on target lane |
| $P_{\text{lc}}$ | Lane-changing probability |
| $a'_p$ | Acceleration rate of MAV in docking mode |
| $v'_{\max}$ | Maximum speed limit of MAV in independent moving and collective moving modes |
| $d_{\text{intra}}$ | Spacing between individual MAV modules within MAV train |
| $v_{\text{det}}^{\text{leader}}$ | Speed of leader module in a MAV-train |
| $P_{\text{d}}$ | MAV module detachment probability |
| $P_{\text{rand}}$ | Random seed $\in (0,1)$ |
| $L_{\max}$ | Maximum module unit within a MAV train |

## REFERENCES

1. Chen, Z., Li, X., and Zhou, X. Operational Design for Shuttle Systems with Modular Vehicles under Oversaturated Traffic: Discrete Modeling Method. *Transportation Research Part B: Methodological*, 2019. 122: 1-19.

2. Chen, Z., Li, X., and Zhou, X. Operational Design for Shuttle Systems with Modular Vehicles under Oversaturated Traffic: Continuous Modeling Method. *Transportation Research Part B: Methodological*, 2020. 132: 76-100.

3. Zhang, Z., Tafreshian, A., and Masoud, N. Modular Transit: Using Autonomy and Modularity to Improve Performance in Public Transportation. *Transportation Research Part E: Logistics and Transportation Review*, 2020. 141: 102033.

4. Wu, J., Kulcsár, B., and Qu, X. A Modular, Adaptive, and Autonomous Transit System (MAATS): An In-motion Transfer Strategy and Performance Evaluation in Urban Grid Transit Networks. *Transportation Research Part A: Policy and Practice*, 2021. 151: 81-98.

5. Chen, Z., and Li, X. Designing Corridor Systems with Modular Autonomous Vehicles Enabling Station-wise Docking: Discrete Modeling Method. *Transportation Research Part E: Logistics and Transportation Review*, 2021. 152: 102388.

6. Chen, Z., Li, X., and Qu, X. A Continuous Model for Designing Corridor Systems with Modular Autonomous Vehicles Enabling Station-wise Docking. *Transportation Science*, 2022. 56(1): 1-30.

7. Lin, J., Nie, Y. M., and Kawamura, K. An Autonomous Modular Mobility Paradigm. *IEEE Intelligent Transportation Systems Magazine*, 2022: 15(1): 378-386.

8. Li, Q., and Li, X. Trajectory Planning for Autonomous Modular Vehicle Docking and Autonomous Vehicle Platooning Operations. *Transportation Research Part E: Logistics and Transportation Review*, 2022. 166: 102886.

9. Li, Q., and Li, X. Trajectory Optimization for Autonomous Modular Vehicle or Platooned Autonomous Vehicle Split Operations. *Transportation Research Part E: Logistics and Transportation Review*, 2023. 176: 103115.

10. Han, Y., Ma, X., Yu, B., Li, Q., Zhang, R., & Li, X. Planning two-dimensional trajectories for modular bus enroute docking. *Transportation Research Part E: Logistics and Transportation Review*, 2024, 192, 103769.

11. Tian, J., Li, G., Treiber, M., Jiang, R., Jia, N., and Ma, S. Cellular Automaton Model Simulating Spatiotemporal Patterns, Phase Transitions and Concave Growth Pattern of Oscillations in Traffic Flow. *Transportation Research Part B: Methodological*, 2016. 93: 560-575.

12. Ye, L., and Yamamoto, T. Modeling connected and autonomous vehicles in heterogeneous traffic flow. *Physica A: Statistical Mechanics and its Applications*, 2018, 490, 269-277.

13. Treiber, M. and Kesting, A. Traffic Flow Dynamics. *Traffic Flow Dynamics: Data, Models and Simulation*, Springer-Verlag Berlin Heidelberg, 2013.